\documentclass[aps,twocolumn,amsmath,letterpaper]{revtex4-1}
\usepackage{amssymb}
\usepackage{graphicx}
\usepackage{array}
\usepackage{hhline}
\usepackage{longtable}
\usepackage{xy}
\usepackage{multirow}

\renewcommand{\thetable}{\Roman{table}} \thetable

\begin{document}
\title{A Hierarchy of Multi-Lane “Entropy Machines” with Unfair Resource Availability}
\author{Ayse Ferhan Yesil*}
\author{M. Cemal Yalabik}
\affiliation{Department of Physics, Bilkent University, Bilkent,
06800, Ankara, Turkey}
\date{\today}
\begin{abstract}
We present a model system for objects, climbers, which have the ability to move as ASEP  particles with non-uniform forward and backward jumping rate dynamics.
These climbers are implemented as Mandal-Quan-Jarzynski [1] machines. Climbers on a number of neighboring columns are considered. The low entropy resource is provided abundantly to the first column, the unused part of this resource is consecutively passed over to objects in neighboring columns. This results in a hierarchy among the lanes. The system displays interesting steady-states. Both open and periodic boundary conditions are considered and results from Monte Carlo simulations of the system will be reported.

\end{abstract}
\maketitle
\def\s{\rule{0in}{0.28in}}

\section{Introduction}
\setlength{\LTcapwidth}{\columnwidth}


Our understanding of the association of physical means of handling information with thermodynamic entropy has evolved due to the seminal work of Landauer \cite{Landauer}.
This work led to the understanding that the entropy produced by generating randomness in some type of physically stored data must be taken into consideration when 
calculating the change in the total entropy associated with the functioning of the system.

Recently, Mandal {\em et al.}~have demonstrated the possibility of a refrigerator running on information\cite{Jar-PNAS,Jar-PRL}. 
The system extracts heat from a low temperature source and places
it at a higher temperature, without utilizing any mechanical energy input. Second law is still satisfied as entropy is produced by irreversibly modifying the contents of
a ``tape'' which is used as a medium to process information associated with the state of the system. The authors allude to the Maxwell Demon for obvious reasons.

In this present work, we study the collective behavior of a number of ``machines'' which run on information. The inter-dependency of the machines is due to a configuration
in which the information output of some machines are fed to other machines lower in a hierarchy of information access. 
Machines in the highest level of hierarchy have access to 
a periodic supply of low-entropy information.
Machines at lower levels of hierarchy have to function with lower efficiency due to the deteriorated information supply.

The model has some features associated with driven diffusive systems, especially with ASEP systems\cite{Krug,Evans}, in terms of its dynamics. 
Although multilane versions of such systems
have been studied\cite{Curatolo}, the emphasis has been on availability of multiple channels for transport, rather than a competition between them. 

Even though it is tempting to associate some particulars of this study
with social structures and their operations, the model obviously is too simplistic even for real physical problems.

\section{The Building Block}

\begin{figure}       
  \begin{center}
     \includegraphics[width=0.5\textwidth]{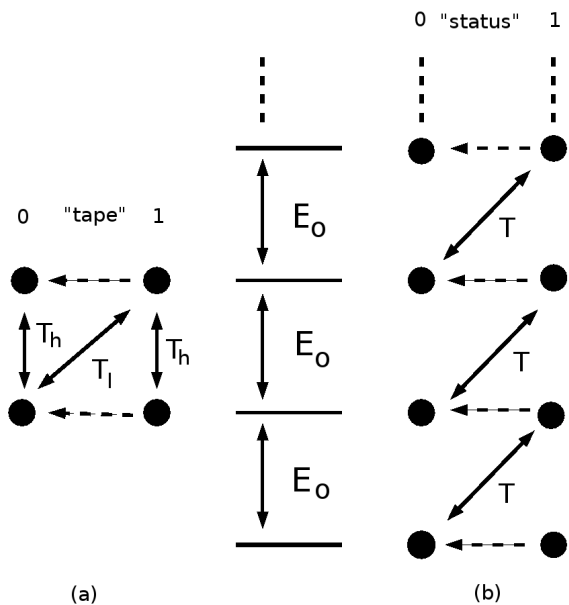}
   \end{center}
  \caption{\label{machines} The refrigerator studied in reference \cite{Jar-PNAS} (a) and the basic building block of the systems analyzed in this work (b). The machines
  reside in one of the states denoted by black circles. Information is accessible as a ``tape'' (or a ``status'' in (b)) bit which is (optimally) refreshed periodically to value zero
  (the dotted arrows). Otherwise, the system may make transitions between two energy levels $E_0$ apart, under the effect of heat baths with temperatures $T_h$ and 
  $T_l$. (A single heat bath at temperature $T$ is present at (b).) With proper choice of parameters, on the average, heat is taken from the low temperature ($T_l$) bath
  and deposited to the bath at the higher temperature $T_h$ by the machine at (a). The machine at (b) will eventually end up at the higher energy state, extracting $E_0$
  from the single heat bath at temperature $T$.}
\end{figure}

Figure \ref{machines} gives a brief description of the refrigerator proposed by Mandal {\em et al.}\cite{Jar-PNAS,Jar-PRL} and 
the machine which is used as a building block of various system configurations studied in this work.

Note that in both machines, transitions may influence a status bit,
which is set to a new value periodically, with period $\tau$. (See references \cite{Jar-PNAS,Jar-PRL}.)
The periodic updating of the status bit should be interpreted as setting the machine status to an ``input'' bit, while the previous status is released as ``output''. (A ``tape''
structure is envisioned which contains these input and output bits.)
Best machine performance will be obtained if this new value is always equal to zero, corresponding to a low entropy. 

The machine in Figure \ref{machines}b for example, will start its cycle at a certain energy level in status zero. It will end up in the higher energy state at the end of the period,
with status one, after the time period $\tau$, with probability $p_{c,u}=\omega_u[1-\exp(-\Omega \tau)]/\Omega$. 
Here, $\Omega$ is the sum of $\omega_u$ for the transition rate from the lower energy level
to the higher one, and $\omega_d$ for the reverse process. The transitions are driven by the heat bath at temperature $T$ so that $\omega_d/\omega_u = \exp(E_0/k_BT)$.
Therefore, energy is extracted from the single heat bath at the rate $\mbox{P}=E_0 p_{c,u} / \tau$, corresponding to an overall entropy reduction rate $E_0 p_{c,u} /T\tau$.
Second law is not violated because conversion of the status bit from zero to one with probability $p_{c,u}$ corresponds to an entropy production rate of 
$-k_B[p_{c,u} \log(p_{c,u}) + (1-p_{c,u})\log(1-p_{c,u})]/\tau$.
%

\begin{figure}       
  \begin{center}
     \includegraphics[width=0.5\textwidth]{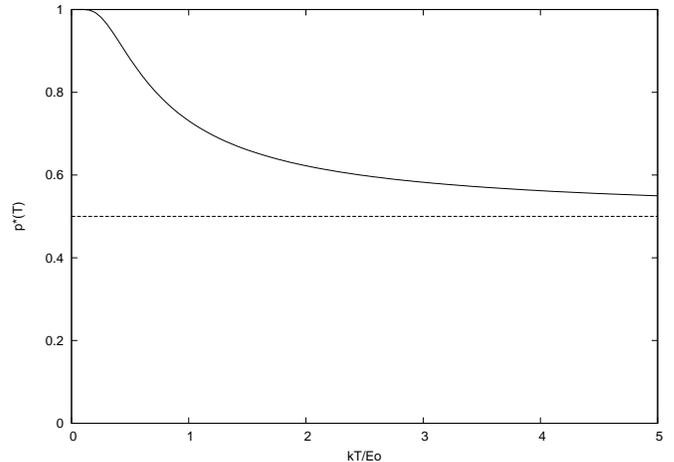}
   \end{center}
  \caption{\label{p_star} The minimum value for $p_0$ as a function of $k_BT/E_0$ in order to
  produce positive amout of work.
  Note that $p_0$ must be close to one in order to extract work from a low temperature bath,
  while for a high temperatue bath, $p_0$ values closer to 0.5 may be used.}
\end{figure}

\begin{figure}       
  \begin{center}
     \includegraphics[width=0.5\textwidth]{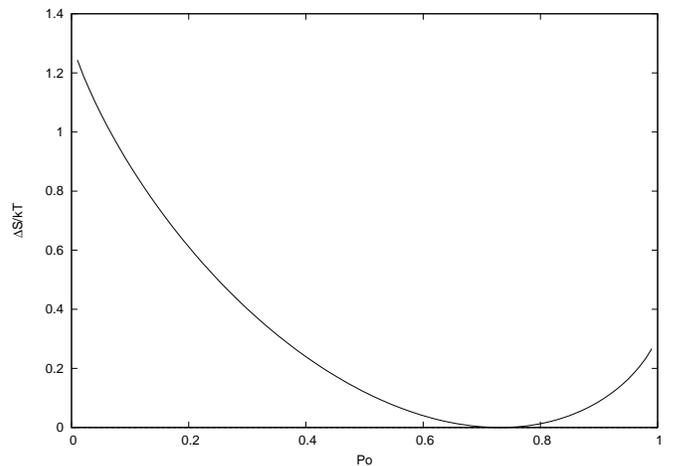}
   \end{center}
  \caption{\label{total_entropy} Total entropy produced by the machine for $\Omega \tau = 3$ and
  $k_BT/E_0 = 1$ yielding a value $p^*(T)\sim 0.73$.}
\end{figure}

We now proceed to study this machine, when it is operating with random input status bits, equal to zero with probability $p_0$. Note that we now allow the possibility that
the energy state of the machine may go down a step within time $\tau$, as its status may be set to one by the input stream. The probability that the energy state of the machine
is lowered by $E_0$ at the end of the period $\tau$ is $p_{c,d} =\omega_d[1-\exp(-\Omega \tau)]/\Omega$. We then have the following probabilities 
just before the next periodic status update:
\begin{equation} \begin{array}{lcl}
    P(\mbox{Higher energy level}) & = & p_0 p_{c,u} \\
    P(\mbox{Lower energy level}) & = &  (1-p_0) p_{c,d} \\
    P(\mbox{Same energy level}) & = & p_0 (1 - p_{c,u}) + (1-p_0)(1-p_{c,d}) \nonumber
\end{array}\end{equation}
$\;$

The rate at which energy is extracted from the heat bath will be $\mbox{P}=sE_0/\tau$ with $s$ the expectation value of the change in the energy ladder step: 
$s=p_0 p_{c,u} - (1-p_0)p_{c,d}$. Note that in order for this quantity to be positive, one needs to have $p_0/(1-p_0) > p_{c,d}/p_{c,u} = \exp(E_0/k_BT)$.
This condition implies 
\[ p_0 > p^*(T) \]
where
\[ p^*(T) = 1/(1+\exp(-E_0/k_BT)).\]
The above relations can then be written in a more compact way using
\[ s = (p_0 - p^*(T))[1-\exp(-\Omega \tau)] . \]
The probability that the machine will release a zero (as the previous status state) at the next periodic status update will be
$p'_0 = p_0(1-p_{c,u}) + (1-p_0) p_{c,d}$
which may also be written as
\[ p'_0 = p_0 + (p^*(T) - p_0)[1-\exp(-\Omega \tau)] .\] 
(Primed variables will be used to identify the quantities associated with the output of the machine.) 
It is apparent that the machine behaves in qualitatively different forms when $p_0 < p^*(T)$ and $p_0 > p^*(T)$. When $p_0 > p^*(T)$,
mechanical work is generated using the energy input from the heat bath. The string of zeros and ones released by the machine then
has a $p'_0$ closer to $p^*(T)$ in comparison to the input string. On the other hand, for $p_0 < p^*(T)$ mechanical work is converted to
heat, but the entropy of the output string is reduced with a higher probability of zeros. The machine then acts as an ``eraser''.
Figure \ref{p_star} shows the variation of $p^*(T)$ as a function of temperature. Extraction of mechanical work from
a lower temperature heat bath requires higher values for $p_0$.

The entropy change associated with the mechanical energy is
\[ S_{mech}/k_B = \frac{E_0}{k_BT}(p^* - p_0)[1-\exp(-\Omega \tau)]\]
while the change in the information entropy is
\[ \Delta S_I = S_I(p'_0) - S_I(p_0)\]
with
\[ S_I(p)/k_B = -p\ln p -(1-p)\ln(1-p).\]
In any case, the overall entropy increases, as expected (Figure \ref{total_entropy}). 

If a second machine
is using the output of the first machine as its input, the corresponding parameter for the work takes a very simple form: 
$s'=p' \Delta - (1-p')\delta = s\exp(-\Omega\tau)$. That is, if the periodic status update time is sufficiently long, the output status bits produced by the first
machine, on the average, is essentially a useless sequence for utilization in another identical machine.

\section{The Model}

\begin{figure}       
  \begin{center}
     \includegraphics[width=0.5\textwidth]{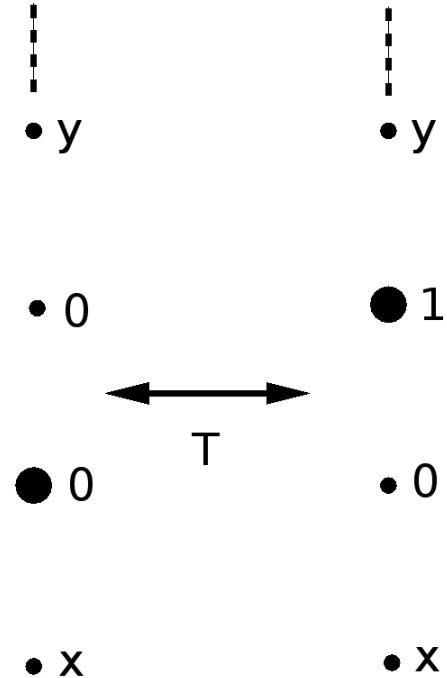}
   \end{center}
  \caption{\label{column} Motion of a machine along a column. Small circles denote the possible positions of a machine, with increasing energy in the vertical direction.
  Larger circle represents the position of the machine. It may move up or down along the energy ladder under the influence of the heat bath at temperature $T$. The numbers
  next to the circles represent the status bits that change during the transition. The $x$ and $y$ variables denote the status bits which remain unchanged.}
\end{figure}

A model is now proposed in which the structure in Figure \ref{machines}b is repeated in neighboring columns. Machines are allowed to climb up (or down) each column, 
height along the column being associated with the energy state of the machine. (We will allow more than one machine to operate on a column.) 
Different status bits are associated with each position of the thus formed lattice. (Initially all status bits are set to one.) A description of the dynamics of a machine
along a column is given in Figure \ref{column}.
Machines may modify the status bits of the sites they move into: A machine with status zero may move up in the lattice, only if the new site is unoccupied and has status zero.
The status of the new site will then be modified as one. Restrictions for going down in energy is the reverse: Machine must start from a site with status one, and may land 
in a new (unoccupied) site with status 0. The status of the original site is returned to zero. At refresh times all status bits are shifted one column to the column on the right.
Fresh status bits (all zeros) are fed from the left to all positions of the leftmost column. 

This structure then puts the machines in the leftmost column in a more advantageous
position: They can only move up in energy, possibly being restricted by the presence of other machines in the same column. This type of motion is a type of TASEP (totally asymmetric
simple exclusion process\cite{Krug,Evans}), with a somewhat complicated jump rate. The status bits drifting towards the right at periodic refresh intervals
provide columns to the right with lower and lower ``quality'' status bits to work with.

We study two types of models: In both of them there are 40 columns. In the first model, we consider a semi-infinite lattice in the energy direction, 
with a lower bound ``ground state'' energy. We start with a completely unoccupied lattice, with an entry rate of $\alpha$ into the lowest energy state if it is unoccupied and
has status zero. (This again provides a hierarchical advantage to columns that are closer to the left boundary.) Note that this system will not reach a steady state, 
and we report the statistical results of a finite time (albeit unfair) ``race'' among the columns.

In the second model we assume periodic boundary conditions in the energy direction. (Step-wise motion in such non-conservative force fields has been very eloquently studied by
Escher\cite{Escher}.) Here, we produce steady state results which indicate that although this race on a ring is still unfair, disadvantaged machines tend to align themselves 
at positions where they avoid the partially depleted outputs of machines to their left, and use the the remnants of the fresh  status bits that were untouched.

\section{Results}

The models described above using a Kinetic Monte Carlo procedure\cite{Kinetic_MC}. The values reported in this section
typically involve the average of 100 independent runs for the semi-infinite system and $10^6$ Monte Carlo
steps for the ring systems. (One Monte Carlo step is defined as the period in which all objects in the
system move once on the average.)

The first column of the first model (on a semi-infinite lattice) is a TASEP system, as mentioned before. Although the infinite size of the lattice does not allow a steady state 
to be reached through a simulation, the current and density near the entry point (the ground state) should reach a steady state. A direct comparison with known results for 
this model is not available, due to the complications introduced to the transition rates through the periodic status update. Nevertheless, one would expect to see the phase
transition (apparent due to the discontinuities in the second derivatives of density and current as a function of entry rate) near a value of the entry rate which produces a
density $\rho\sim 0.5$. Studies on finite size TASEP 
systems with open boundaries yield a ``bulk'' density which changes as a function of $\alpha$ until a critical value is reached ($\alpha_c =0.5$ for the 
simple system in \cite{Evans}). For values of $\alpha > \alpha_c$ the bulk density stays fixed at 0.5. (Actually, in this regime, the ``bulk'' region is formed
by a linear change in density, averaging to the value 0.5, between the entry and exit boundary regions.) 

\begin{figure}       
  \begin{center}
     \includegraphics[width=0.5\textwidth]{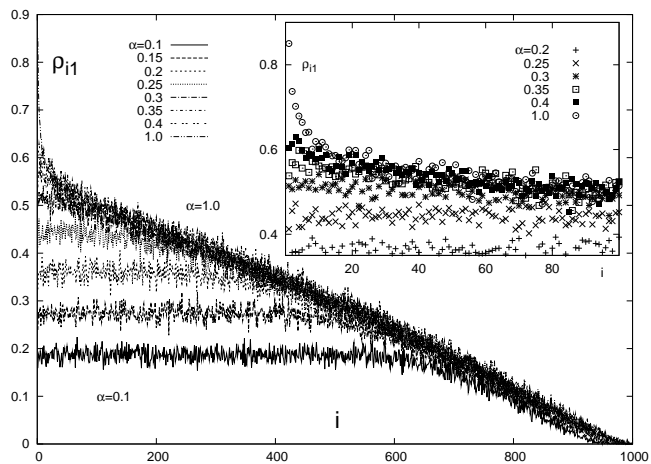}
   \end{center}
  \caption{\label{dens_plot} Density profile along the first column, for various values of the entry rate $\alpha$. The inset shows the detail near the entry boundary.
  Note that apart from a small boundary region, the density profile becomes undiscernable for $\alpha \ge 0.35$.}
\end{figure}

Figure \ref{dens_plot} shows the density profile along the first column, for various values of the entry rate $\alpha$. The density of machines away from the entry boundary
seem to have a universal profile, independent of the value of $\alpha$. This profile is formed by the pioneer machines which are ``filling up'' empty space in front of them.
The density behind this front depends on the value of $\alpha$: For values of $\alpha$ greater than about 0.35, the density front reaches very near the entry boundary.
For smaller values of $\alpha$, the density profile reaches the entry boundary with a value which depends on $\alpha$, with no appreciable boundary effect. The critical density value 
corresponding to  the change of behavior is again 0.5. 


\begin{figure}       
  \begin{center}
     \includegraphics[width=0.5\textwidth]{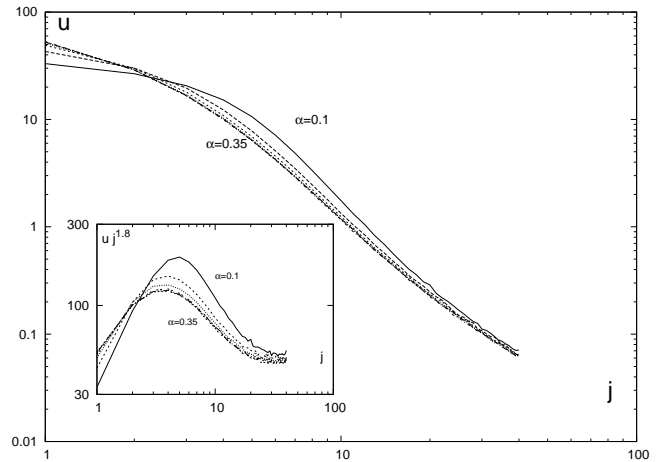}
   \end{center}
  \caption{\label{util_plot} Total energy produced in the $j$th column (in units of $E_0$) with $u_j = \sum_i i\rho_{ij}$ corresponding to the distribution in Figure \ref{dens_plot}.
  The inset is a plot of $j^{1.8} u_j$ to display the power law behavior. The plots are for $\alpha =$ 0.1, 0.15, 0.2, 0.25, 0.3 and 0.35. Note that curves for $\alpha > 0.2$ are 
  undiscernable.}
\end{figure}

The current along the columns is a measure of of the rate at which energy is being absorbed from the heat bath. That in turn is related to the supply of low entropy
status bits to the machines. The total current at the entry boundary, corresponding to the energy production rate of machines on 40 columns (each feeding on the remnants 
of the entropy resource left from the previous column), adds up to what is available at the leftmost column. Indeed, we find that the total entry rate of machines to the 
system is 
independent of the value of $\alpha$. How that resource utilization rate is distributed among the columns does of course depend on $\alpha$, and is
shown in Figure \ref{util_plot}. Note that smaller values of the entry rate correspond to more a equitable (among the hierarchies) distribution for the utilization of 
the low entropy resource.

Authors acknowledge support from Turkish Academy of Sciences.

\end{document}